\documentclass[preprint,12pt]{elsarticle}

\usepackage{color}
\usepackage[cmex10]{amsmath}
\usepackage{float}
\usepackage{subfig}
\usepackage{algorithmic}
\usepackage{array}
\usepackage{url}
\usepackage{multirow}
\usepackage{algorithm}
\usepackage{algorithmic}
\usepackage{amsmath}
\usepackage{amssymb}
\usepackage{array}
\usepackage{longtable}
\usepackage{rotating}
\usepackage{etoolbox}
\usepackage{graphicx}  
\usepackage{graphicx}
\usepackage{float} 
\usepackage{subfloat}
\usepackage{makecell}
\usepackage{textcomp}
\usepackage{threeparttable}
\usepackage{booktabs}
\usepackage{setspace}
\usepackage{hyperref}
\hypersetup{hidelinks}

\begin{document}

\begin{frontmatter}

\title{A generalized motif-based Na\"ive Bayes model for sign prediction in complex networks} 

\author[label1,label2]{Yijun Ran}
\ead{ranyij288@gmail.com}
		
\author[label3]{Si-Yuan Liu}
\ead{liusiyuan311@163.com}

\author[label3]{Junjie Huang}
\ead{junjiehuang@swu.edu.cn}
		
\author[label4,label3]{Tao Jia}
\ead{tjia@swu.edu.cn}
		
\author[label2]{Xiao-Ke Xu \corref{cor1}}
\ead{xuxiaoke@foxmail.com}

\cortext[cor1]{Corresponding author.}
\address[label1]{School of Big Data and Computer Science, Guizhou Normal University, Guiyang 550025, People's Republic of China}
\address[label2]{Center for Computational Communication Research, Beijing Normal University Zhuhai, 519087, People's Republic of China}
\address[label3]{College of Computer and Information Science, Southwest University, Chongqing 400037, People's Republic of China}
\address[label4]{College of Computer and Information Science, Chongqing Normal University, Chongqing 401331, People's Republic of China}

\begin{abstract}

Signed networks, encoding both positive and negative interactions, are essential for modeling complex systems in social and financial domains. Sign prediction, which infers the sign of a target link, has wide-ranging practical applications. Traditional motif-based Na\"ive Bayes models assume that all neighboring nodes contribute equally to a target link's sign, overlooking the heterogeneous influence among neighbors and potentially limiting performance. To address this, we propose a generalizable sign prediction framework that explicitly models the heterogeneity. Specifically, we design two role functions to quantify the differentiated influence of neighboring nodes. We further extend this approach from a single motif to multiple motifs via two strategies. The generalized multiple motifs-based Na\"ive Bayes model linearly combines information from diverse motifs, while the Feature-driven Generalized Motif-based Na\"ive Bayes (FGMNB) model integrates high-dimensional motif features using machine learning. Extensive experiments on four real-world signed networks show that FGMNB consistently outperforms five state-of-the-art embedding-based baselines on three of these networks. Moreover, we observe that the most predictive motif structures differ across datasets, highlighting the importance of local structural patterns and offering valuable insights for motif-based feature engineering. Our framework provides an effective and theoretically grounded solution to sign prediction, with practical implications for enhancing trust and security in online platforms.
\end{abstract}

\begin{keyword}
Sign prediction \sep Network motifs \sep Na\"ive Bayes model \sep Signed networks
\end{keyword}

\end{frontmatter}

\section{Introduction}
Social media has transformed how individuals connect and interact through activities like following, commenting, and retweeting, making social networks central to modern communication by linking individuals and organizations through diverse relationships \citep{allen2024quantifying,park2023invasion,ogushi2017enhanced,xie2025towards}. These networks capture both positive interactions, such as friendship and trust, and negative ones, like distrust and conflict \citep{leskovec2010signed,tang2016survey,leskovec2010predicting}. Understanding this duality is crucial for comprehensively analyzing social dynamics. However, the signs of only some links are available in many cases, while others are missing due to measurement gaps, lack of recording, or unreliable data \citep{kirkley2019balance,ran2022predicting}. Hence, sign prediction, which aims to infer the sign of these relationships, provides valuable insights into the balance of cooperation and conflict within networks, with applications in recommendation systems \citep{liu2021effective,seo2022siren}, political analysis \citep{keuchenius2021important,godziszewski2024adversarial}, and cybersecurity \citep{taddeo2019trusting,liu2024event}.

Sign prediction, analogous to link prediction, aims to infer the sign of a targeted link in a signed network based on the known signs of other links \citep{kirkley2019balance,li2020learning,lin2022status}. Sign prediction is not only a predictive task but also a means of uncovering the latent social, economic, and political dynamics embedded in signed networks. By combining predictive accuracy with theoretical insights from structural balance theory, it advances computational methods while deepening our understanding of complex relational systems. Methods for sign prediction can be broadly categorized into feature-based and network-embedding approaches \citep{shen2018deep,fang2023signed}. Network-embedding methods map signed networks into low-dimensional vector spaces, enabling the learning of node representations that can be used for sign prediction \citep{lin2022status,shen2018deep,hou2020network}. These methods often draw on structural theories such as balance and status theory to guide the embedding process. For example, the Signed Graph Convolutional Network (SGCN) leverages balance theory by employing graph convolutional layers to embed nodes in undirected signed networks \citep{derr2018signed}. Extending this approach to directed networks, the Signed Directed Graph Neural Network (SDGNN) introduces enhanced attention mechanisms and a tailored loss function to simultaneously reconstruct link signs, directions, and triadic patterns, effectively addressing the complexities of directed signed relationships \citep{huang2021sdgnn}. 

Feature-based methods primarily leverage network topology and social psychology principles to extract features for predicting the sign of links. Among these, All23 is a well-known method that extracts 23 features, including signed network degree and triadic features, for sign prediction \citep{leskovec2010signed,leskovec2010predicting}. These features are grounded in structural balance theory and status theory. Structural balance theory posits that certain triadic configurations are more likely in balanced networks, reflecting harmony or conflict among nodes. Status theory, in turn, effectively explains local patterns of signed links and captures richer aspects of user behavior, offering a nuanced perspective on interaction dynamics. 

Network motifs, as fundamental structural and functional modules in real-world networks, have been widely used in sign prediction as part of feature-based methods \citep{papaoikonomou2014predicting,song2015link,park2021motif}. Most existing approaches involve counting the number of motifs formed with the target link and using this count as a feature input to machine learning algorithms \citep{leskovec2010predicting,chiang2011exploiting,zhang2013potential}. However, these methods for sign prediction often fail to differentiate the roles of individual nodes within a motif. To address this limitation, Liu \textit{et al.} propose the Single Motif-based Na\"ive Bayes (SMNB) model, which not only elucidates the prediction mechanism of motif-based methods but also quantifies the contribution of each neighboring node to sign prediction \citep{liu2020sign}. While the SMNB model effectively evaluates the influence of neighboring nodes in predicting the sign of a target link, it assumes that all links sharing the same neighbor have identical probabilities of being positive or negative. This assumption limits the model’s granularity and reduces its adaptability to complex network structures, where a single neighbor may contribute differently to the sign of different target links. Although SMNB has made significant progress, the heterogeneous influence of a neighboring node on the sign of links remains insufficiently explored.

To address the above limitation, we consider the heterogeneous influence of a neighboring node on the probability of a link being positive or negative. Here, we propose two enhanced approaches: the Generalized Single Motif-based Na\"ive Bayes model based on Common Link (GSMNB-CL) and the Generalized Single Motif-based Na\"ive Bayes model based on Common Node (GSMNB-CN). Specifically, GSMNB-CL focuses on motifs that include a neighboring node and at least one common link connecting a node in the target link to the neighbor. In contrast, GSMNB-CN examines motifs that include only the neighboring node, explicitly excluding any common links between the target link and the neighbor. Furthermore, we explore two strategies for quantifying the predictive performance of multiple motifs in sign prediction. The first strategy extends GSMNB-CL using Na\"ive Bayes theory to accommodate multiple motifs, resulting in the Generalized Multiple Motifs-based Na\"ive Bayes (GMMNB) model. The second strategy employs a machine-learning framework, referred to as FGMNB, where the scores of multiple motifs computed by GSMNB-CL are combined into a multi-dimensional feature vector and input into a machine-learning algorithm to assess their collective predictive impact. The main contributions of this work can be summarized as follows:

\begin{enumerate}
	\item To overcome the limitation of traditional motif-based Na\"ive Bayes models, we propose a generalized motif-based Na\"ive Bayes model that accounts for the heterogeneous influence of a neighboring node on the sign of a target link.
	
	\item We introduce two generalized single motif-based Na\"ive Bayes models, GSMNB-CL and GSMNB-CN, designed to capture the heterogeneous influence of a neighboring node. Experimental results on real-world datasets demonstrate that GSMNB-CL consistently outperforms SMNB across all evaluation metrics.
	
	\item To extend the single-motif methods to multiple motifs, we propose two complementary strategies: GMMNB and FGMNB. The experiments demonstrate that FGMNB outperforms five state-of-the-art embedding-based baselines on three of the four real-world signed networks, underscoring the effectiveness of our approach for sign prediction.
	
\end{enumerate}

This study is organized as follows. In Section \ref{section1}, we present a comprehensive review of related work. Section \ref{section2} defines the problem of sign prediction. Section \ref{section3} presents the mathematical formulations of the proposed framework. In Section \ref{section4}, we evaluate the performance of the proposed framework on sign prediction tasks. Finally, Section \ref{section5} provides the conclusions and prospects of this study.

\section{Related Work}
\label{section1}
Sign prediction aims to determine the sign of target links in a signed network based on the known signs of other links. Existing approaches can be broadly classified into feature-based and network-embedding methods. Below, we provide a detailed overview of key studies employing these techniques.

\subsection{Network-embedding methods}
Network embedding aims to learn low-dimensional representations for nodes in a network, which can then be applied to various downstream tasks \citep{fang2023signed,huang2019signed,liang2025line}. Signed network embedding extends this concept to signed networks, capturing both positive and negative relationships. For example, Yuan \textit{et al.} \citep{yuan2017sne} introduce a log-bilinear model for signed network representation, while SiNE \citep{wang2017signed} designs an objective function guided by social theories to enhance the learning process. To preserve structural balance, Shen \textit{et al.} propose a deep network embedding model that employs a semi-supervised stacked autoencoder for sign prediction \citep{shen2018deep}. Chen \textit{et al.} develop the Beside method, which effectively incorporates both triangular and bridge structures to improve sign prediction performance \citep{chen2018bridge}. Javari \textit{et al.} introduced ROSE, a novel transformation-based embedding framework that converts signed networks into unsigned bipartite networks with role-nodes for embedding \citep{javari2020rose}. SGCN redefines information aggregation and propagation in undirected signed networks based on balance theory \citep{derr2018signed}. Huang \textit{et al.}  propose the Signed Graph Attention Network (SiGAT), integrating Graph Attention Networks (GATs) into directed signed networks and designing a motif-based graph neural network guided by social theories \citep{huang2019signed}. Recently, the SELO model adopts a subgraph encoding approach to learn link embeddings through linear optimization, specifically targeting signed and directed networks \citep{fang2023signed}.

\subsection{Feature-based methods}
Feature-based methods usually extract information from network topology, node attributes, and connection patterns to design similarity models for sign prediction. For example, DuBois \textit{et al.} incorporate additional features such as gender, interests, and location in social networks to enhance predictive performance \citep{dubois2011predicting}. Leskovec \textit{et al.} propose an algorithm that leverages structural balance theory and local network features for sign prediction \citep{leskovec2010predicting}. Chiang \textit{et al.} construct feature sets by extracting k-hop rings based on the Katz index and apply logistic regression for sign prediction \citep{chiang2011exploiting}. Wang \textit{et al.} introduce a low-rank matrix factorization framework to estimate user states \citep{wang2015modeling}. Liu \textit{et al.} develop a method that integrates structural balance theory with path-based similarity to effectively capture the influence of paths and attributes on node similarity \citep{liu2017link}. Khodadadi \textit{et al.} propose a sign prediction algorithm based on closed triple microstructures, which efficiently infers unknown relation types while maintaining low computational complexity \citep{khodadadi2017sign}. Furthermore, Liu \textit{et al.}  introduce the SMNB model, which not only explains the predictive mechanism of motif-based methods but also quantifies the contribution of each neighboring node to sign prediction \citep{liu2020sign}.

Network motif features are widely used in feature-based methods for sign prediction. However, most existing approaches utilize only motif counts as predictive features. To overcome this limitation, Liu \textit{et al.} introduced the SMNB model based on Bayesian theory \citep{liu2020sign}. While effective, the SMNB model assumes that all links sharing the same neighbor have identical probabilities of being positive or negative. To address this constraint, we here propose a generalizable approach that captures the heterogeneous influence of neighboring nodes on the sign of a target link.

\section{Problem definition}
\label{section2}
Sign prediction can be considered an analogous problem to link prediction in complex networks \citep{ran2024maximum,lu2011link,shang2017link}. While link prediction seeks to uncover latent relationships that exist but are not explicitly recorded, sign prediction focuses on estimating the sentiment or polarity of relationships between individuals \citep{leskovec2010predicting,kirkley2019balance}. This task becomes particularly critical when some link signs in the network are missing, necessitating their prediction based on the known signs and the structural properties of the network.

Here, we adopt the most common settings of the sign prediction problem \citep{leskovec2010predicting,liu2020sign,javari2014cluster,kumar2016edge}. Assume an undirected signed network $G(N, L, W)$, where $N$ is the set of nodes, $L$ is the set of links, and $W$ is the set of signs associated with the links in $L$. A node in network $G$ cannot connect to itself (no self-loops) nor share more than one link with another node (no repeated links). Let $+_{AB}$ denote a positive sign for the link $(A, B)$, and $-_{AB}$ represent a negative sign for the link $(A, B)$.

\section{Methodology}
\label{section3}

\subsection{Network motifs}
\label{section31}
Network motifs are recurring, significant subgraphs or patterns within a larger network that occur more frequently than would be expected by chance \citep{milo2002network,alon2007network}. These motifs capture local structural features that may have functional or dynamic significance in the network. In the context of network analysis, motifs are used to identify common configurations that contribute to the overall behavior and organization of the system.
\begin{figure}[htbp]
	\centering
	\includegraphics[width=1.0\textwidth]{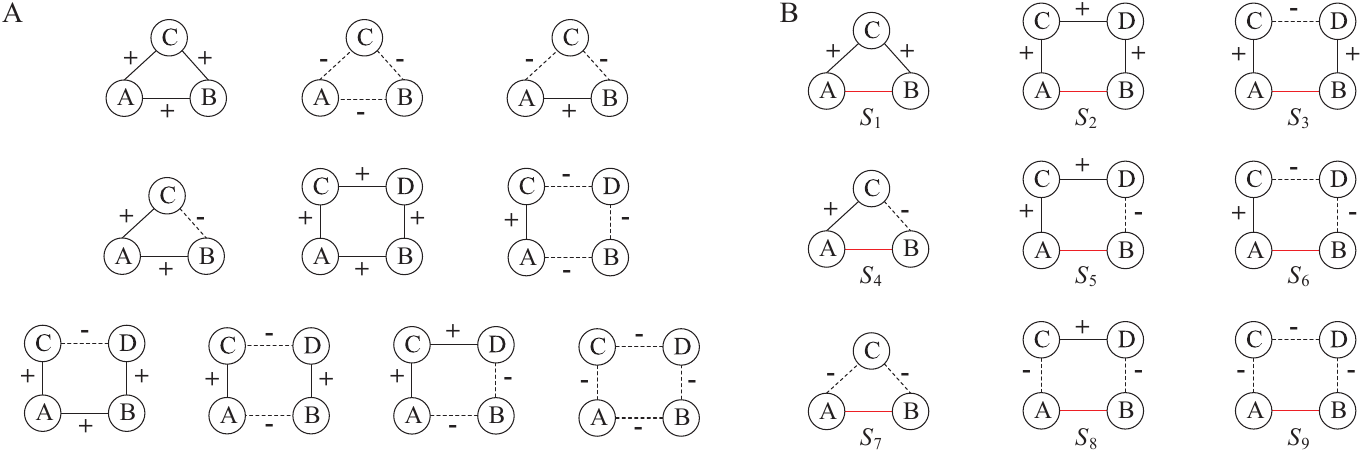}
	\caption{The ten network motifs and nine different predictors. A) We consider ten unique 3-node and 4-node motifs in an undirected signed network. B) The nine predictors are derived from distinct 3-node and 4-node motifs. Each predictor $S_\text{i}$ includes a red line representing the target link whose sign needs to be predicted. For 3-node predictors $S_\text{i}$, we quantify the heterogeneous influence of the neighboring node $C$ on the positive or negative sign of link $(A, B)$. For 4-node predictors $S_\text{i}$, we treat the neighboring link $(C, D)$ as a single entity and evaluate the heterogeneous influence of the neighboring link $(C, D)$ on the sign of link $(A, B)$.}
	\label{Fig:hm}
\end{figure}

In this work, we focus on undirected signed networks, which can generate ten unique 3-node and 4-node motifs. While higher-order motifs may provide more detailed features for prediction tasks, the accuracy improvement tends to diminish for motifs beyond four nodes compared to 4-node motifs \citep{chiang2011exploiting}. Consequently, we restrict our analysis to 3-node and 4-node motifs for sign prediction. Fig. \ref{Fig:hm}A shows all 3-node and 4-node signed motifs in an undirected signed network. Here, the black solid line represents a positive link (+), and the black dashed line indicates a negative one (-). In constructing the predictors in Fig. \ref{Fig:hm}B, we assume that the sign of one link in each motif is unknown and must be predicted. Based on the possible sign combinations of the remaining links, we obtain a total of nine predictors in Fig. \ref{Fig:hm}B. For example, a 3-node motif can form four possible link-sign combinations in the network: +{}+{}+, +{}+{}-, +{}-{}-, and -{}-{}-. However, in the predictors of Fig. \ref{Fig:hm}B, with one link assumed to have an unknown sign, only the remaining two links are considered, yielding three combinations: +{}+, +{}-, and -{}-. Thus, three predictors of 3-node motifs are constructed in this case ($S_\text{1}$, $S_\text{4}$, $S_\text{7}$ in Fig. \ref{Fig:hm}B). By examining the sign combinations of the remaining links in the motif, we derive nine distinct motif predictors (Fig. \ref{Fig:hm}B). Each predictor $S_\text{i}$ corresponds to a specific motif configuration that includes the target link, encapsulating the structural and signed context necessary for effective sign prediction.

To preserve the structural patterns associated with each predictor $S_\text{i}$ in the original network, we extract motif features for each target link by removing only the signs of links in the testing set. Removing the signs of all links in both the training and testing sets prior to extracting motif features would disrupt the predictive patterns associated with the links being evaluated. This selective approach ensures that the integrity of the motif predictors remains intact while maintaining consistency between the training and testing processes.
\begin{figure}[htbp]
	\centering
	\includegraphics[width=0.4\textwidth]{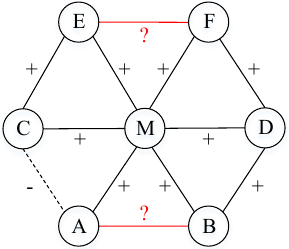}
	\caption{An example demonstrating the different contributions of node $M$ to the probability of links $(A, B)$ and $(E, F)$ having positive or negative signs. For predictor $S_\text{1}$, there are four predictors, $S_\text{1}(A, M, C)$, $S_\text{1} (B, M, D)$, $S_\text{1}(C, M, E)$, and $S_\text{1} (D, M, F)$, formed with the node $M$. For the target link $(A, B)$, links $(A, M)$ and $(B, M)$ appear in $S_\text{1}(A, M, C)$ and $S_\text{1} (B, M, D)$, respectively, however $S_\text{1}(D, M, F)$ and $S_\text{1}(C, M, E)$ only involve the node $M$ and exclude the links $(A, M)$ and $(B, M)$. For the target link (E, F), links $(E, M)$ and $(F, M)$ appear in $S_\text{1}(C, M, E)$ and $S_\text{1} (D, M, F)$, respectively, while $S_\text{1}(A, M, C)$ and $S_\text{1}(B, M, D)$ only involve the node $M$ and exclude the links $(E, M)$ and $(F, M)$.}
	\label{Fig:diff}
\end{figure}

\subsection{Single motif-based Na\"ive Bayes model}
\label{section32}
To distinguish the role of each neighboring node connected to the link whose sign needs to be predicted, Liu et al. proposed the SMNB method \citep{liu2020sign}, which introduces a role function to quantify the role of each neighboring node (for 3-node motifs in Fig. \ref{Fig:hm}B) or neighboring link (for 4-node motifs in Fig. \ref{Fig:hm}B) connected to the target link. In SMNB method, for a given node $M$, the probability $P(+_{AB}|M)$ represents the likelihood that the sign of the link, which forms the predictor $S_\text{i}$ with node $M$, is positive. It can be expressed as
\begin{equation}
P(+_{AB}|M) = \frac{N_{+_{S_\text{i}(M)}}}{N_{+_{S_\text{i}(M)}} + N_{-_{S_\text{i}(M)}}},
\label{eqpos}
\end{equation}
where $N_{+_{S_\text{i}(M)}}$ and $N_{-_{S_\text{i}(M)}}$ denote the number of positive and negative links in the set of links that form the predictor $S_\text{i}$ with common neighbor $M$, respectively. Similarly, $P(-_{AB}|M)$ represents the likelihood that the sign of the link, which forms the predictor $S_\text{i}$ with node $M$, is negative. It can be expressed as
\begin{equation}
P(-_{AB}|M) = 1- P(+_{AB}|M) = \frac{N_{-_{S_\text{i}(M)}}}{N_{+_{S_\text{i}(M)}} + N_{-_{S_\text{i}(M)}}}.
\label{eqneg}
\end{equation}

The SMNB method effectively distinguishes the roles of neighboring nodes in predicting the sign of a target link. However, it assigns the same probability for a positive or negative sign to all links sharing the same neighbor $M$. Take the extraction of predictor $S_\text{1}$ as an example. There are four types of predictors $S_\text{1}$ based on the common neighbor $M$: three are formed by positive links ($(B, D)$, $(D, F)$, and $(C, E)$) with node $M$, and one by a negative link ($(A, C)$) with node $M$ (Fig. \ref{Fig:diff}). According to Eq. \ref{eqpos}, $P(+_{AB}|M) = P(+_{EF}|M)=3/4$, indicating that the probability of links $(A, B)$ and $(E, F)$ being positive is the same under the influence of node $M$. Similarly, according to Eq. \ref{eqneg}, $P(-_{AB}|M) = P(-_{EF}|M)=1/4$, meaning that the probability of both links $(A, B)$ and $(E, F)$ being negative is also the same under the influence of node $M$.

In fact, when predicting the sign of link $(A, B)$, the predictors $S_\text{1}(A, M, C)$ and $S_\text{1} (B, M, D)$ include at least one link from the predictor $S_\text{1}(A, M, B)$ (with links $(A, M)$ and $(B, M)$ appearing in $S_\text{1}(A, M, C)$ and $S_\text{1} (B, M, D)$, respectively), while $S_\text{1}(D, M, F)$ and $S_\text{1}(C, M, E)$ only involve the common neighbor $M$ (Fig. \ref{Fig:diff}). Likewise, when predicting the sign of link $(E, F)$, the predictors $S_\text{1}(C, M, E)$ and $S_\text{1} (D, M, F)$ include at least one link from the predictor $S_\text{1}(E, M, F)$ (with links $(E, M)$ and $(F, M)$ appearing in $S_\text{1}(C, M, E)$ and $S_\text{1} (D, M, F)$, respectively), while $S_\text{1}(A, M, C)$ and $S_\text{1}(B, M, D)$ only involve the common neighbor $M$ (Fig. \ref{Fig:diff}). This suggests that the contribution of $M$ to the probability of links $(A, B)$ and $(E, F)$ having a positive or negative sign should differ.

\begin{figure}[htbp]
	\centering
	\includegraphics[width=0.8\textwidth]{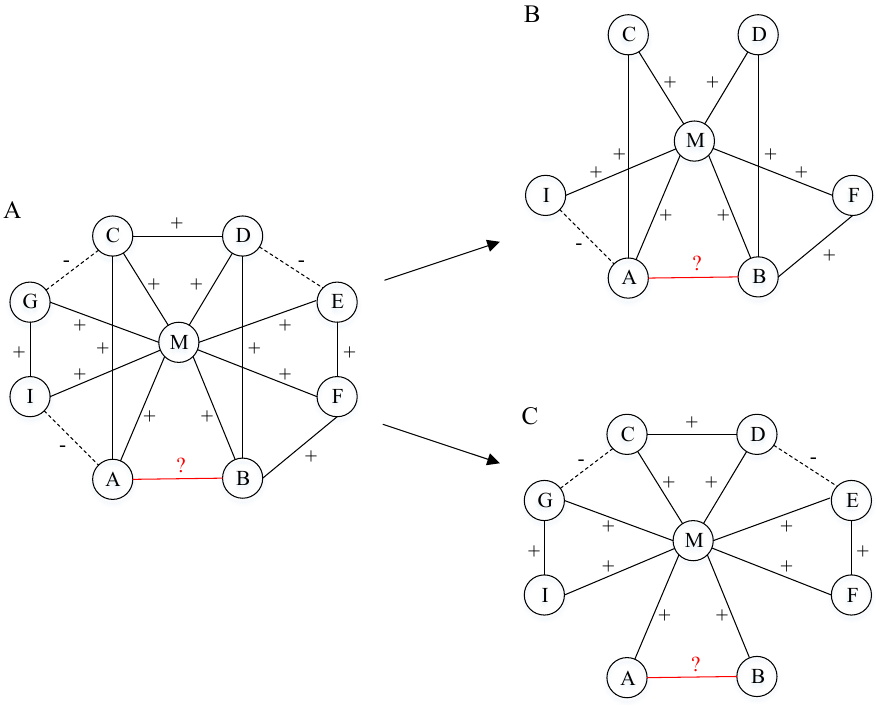}
	\caption{The complete description of two role functions. A) All possible predictors $S_\text{1}$ formed with the node $M$. These predictors can be utilized in two ways. B) One is to consider the predictors $S_\text{1}$ formed with the node $M$ in which each predictor $S_\text{1}$ includes at least one of the links $(A, M)$ and $(B, M)$. C) The other is to consider the predictors $S_\text{1}$ formed with the node $M$ where each predictor includes only node $M$ and excludes the links $(A, M)$ and $(B, M)$.}
	\label{Fig:func}
\end{figure}

\subsection{Generalized single motif-based Na\"ive Bayes model}
\label{section33}
Motivated by the observations in Section \ref{section32}, we here introduce two role functions to quantify the heterogeneous influence of a neighboring node. To predict the sign of link $(A, B)$ (Fig. \ref{Fig:func}A), the following two components can determine the probability of link $(A, B)$ having a positive or negative sign: the likelihood of predictor $S_\text{i}$ that includes at least one link from $(A, M)$ or $(B, M)$, and the likelihood of predictor $S_\text{i}$ that includes only the common neighbor $M$ and excludes the links $(A, M)$ and $(B, M)$. Taking the extraction of predictor $S_\text{1}$ as an example, there are nine types of predictors $S_\text{1}$ formed with common neighbor $M$ (Fig. \ref{Fig:func}A). Among these, four predictors include at least one link from $(A, M)$ or $(B, M)$ (Fig. \ref{Fig:func}B), and five predictors include only the common neighbor $M$ and exclude the links $(A, M)$ and $(B, M)$ (Fig. \ref{Fig:func}C). To quantify the two role functions, we propose the Generalized Single Motif-based Na\"ive Bayes model based on Common Link (GSMNB-CL) and the Generalized Single Motif-based Na\"ive Bayes model based on Common Node (GSMNB-CN). These are two mechanisms for assessing the heterogeneous influence of a neighboring node on the probability of a link's sign being positive or negative.

For a given undirected signed network $G(N, L, W)$, assume we need to predict the sign of a link $(A, B)$. Based on Bayesian theory \citep{bernardo2009bayesian,berger2013statistical}, the posterior probability of $(A, B)$ being positive or negative can be calculated as
\begin{equation}
P(+_{AB}|S_\text{i}(A,B)) = \frac{P(+_{AB})P(S_\text{i}(A,B)|+_{AB})}{P(S_\text{i}(A,B))},
\label{equation3}
\end{equation}
\begin{equation}
P(-_{AB}|S_\text{i}(A,B)) = \frac{P(-_{AB})P(S_\text{i}(A,B)|-_{AB})}{P(S_\text{i}(A,B))},
\label{equation4}
\end{equation}
where $S_\text{i}(A, B)$ denotes a set of nodes that form predictor $S_\text{i}$ in conjunction with nodes $A$ and $B$. For the target link $(A, B)$, the sign is determined using Maximum A Posteriori estimation \citep{greig1989exact} by computing the ratio of these two posterior probabilities. Assume that the contribution of each node $M$ in $S_\text{i}(A, B)$ to the probability of link $(A, B)$ having a positive or negative sign is independent. Hence, the likelihood score for the sign of target link $(A, B)$ can be calculated as
\begin{align}
r_{AB}  &= \frac{P(+_{AB}|S_\text{i}(A,B))}{P(-_{AB}|S_\text{i}(A,B))} \nonumber \\
            &= \frac{P(+_{AB})}{P(-_{AB})} \frac{P(S_\text{i}(A,B)|+_{AB})}{P(S_\text{i}(A,B)|-_{AB})} \nonumber \\
            &= \frac{P(+_{AB})}{P(-_{AB})} \prod_{M \in S_\text{i}(A,B)}\frac{P(M|+_{AB})}{P(M|-_{AB})} \nonumber \\
            &= \frac{P(+_{AB})}{P(-_{AB})} \prod_{M \in S_\text{i}(A,B)} \frac{P(-_{AB})}{P(+_{AB})} \frac{P(+_{AB}|M)}{P(-_{AB}|M)}.
\label{equation5}
\end{align}
Here, $P(+_{AB})$ represents the prior probability that nodes $A$ and $B$ are linked by a positive sign, and $P(-_{AB})$ denotes the prior probability of a negative sign between them. Hence, $P(+_{AB})$ and $P(-_{AB})$ can be computed as
\begin{equation}
P(+_{AB}) = \frac{p|L|}{|L|}=f_+,
\label{equation6}
\end{equation}
\begin{equation}
P(-_{AB}) = \frac{n|L|}{|L|}=f_-.
\label{equation7}
\end{equation}
Let $a = \frac{P(-_{AB})}{P(+_{AB})}=f_-/f_+$, Eq. (\ref{equation5}) can be rewritten as
\begin{equation}
r_{AB} = a^{-1} \prod_{M \in S_\text{i}(A,B)}a {R_{S_\text{i}(M)}},
\label{equation8}
\end{equation}
where $R_{S_\text{i}(M)}= \frac{P(+_{AB}|M)}{P(-_{AB}|M)}$. Since $a$ is a constant for a given network, the factor $a^{-1}$ can be omitted. Taking the logarithm of Eq. (\ref{equation8}), the new likelihood score for the sign of target link $(A, B)$ can be calculated as
\begin{align}
\tilde{r}_{AB} &= \log( \prod_{M \in S_\text{i}(A,B)}a {R_{S_\text{i}(M)}}) \nonumber \\
                     &= \sum_{M \in S_\text{i}(A,B)}\log(a{R_{S_\text{i}(M)}}) \nonumber \\
                     &= |S_\text{i}(A,B)|\log{a}  + \sum_{M \in S_\text{i}(A,B)}\log{R_{S_\text{i}(M)}},
\label{equation9}
\end{align}
where $|S_\text{i}(A, B)|$ denotes the number of the set of nodes that form predictor $S_\text{i}$ in conjunction with nodes $A$ and $B$.

For the GSMNB-CL method, $R_{S_\text{i}(M)}$ in Eq. (\ref{equation9}) can be estimated as
\begin{align}
R_{S_\text{i}(M)}^{\text{GSMNB-CL}} &= \frac{P(+_{AB}|M)}{P(-_{AB}|M)}= \frac{\frac{CLN_{+_{S_\text{i}(M)}}}{CLN_{+_{S_\text{i}(M)}} + CLN_{-_{S_\text{i}(M)}}}}{\frac{CLN_{-_{S_\text{i}(M)}}}{CLN_{+_{S_\text{i}(M)}} + CLN_{-_{S_\text{i}(M)}}}} \nonumber \\
                                                           &= \frac{CLN_{+_{S_\text{i}(M)}}}{CLN_{-_{S_\text{i}(M)}}},
\label{equation10}
\end{align}
where $CLN_{+_{S_\text{i}(M)}}$ represents the number of predictors $S_\text{i}$ formed by a positive link with node $M$ in which each predictor $S_\text{i}$ includes at least one of the links $(A, M)$ and $(B, M)$. Similarly, $CLN_{-_{S_\text{i}(M)}}$ denotes the number of predictors $S_\text{i}$ formed by a negative link with node $M$ where each predictor $S_\text{i}$ includes at least one of the links $(A, M)$ and $(B, M)$. To avoid the expression becoming meaningless due to a zero denominator, 1 is added to both the numerator and denominator in Eq. (\ref{equation10}). Hence, Eq. (\ref{equation10}) can be defined as
\begin{equation}
\tilde{R}_{S_\text{i}(M)}^{\text{GSMNB-CL}} = \frac{CLN_{+_{S_\text{i}(M)}}+1}{CLN_{-_{S_\text{i}(M)}}+1}.
\label{equation11}
\end{equation}

For the GSMNB-CN method, $R_{S_\text{i}(M)}$ in Eq. (\ref{equation9}) can be estimated as
\begin{align}
R_{S_\text{i}(M)}^{\text{GSMNB-CN}} &= \frac{P(+_{AB}|M)}{P(-_{AB}|M)}= \frac{\frac{CNN_{+_{S_\text{i}(M)}}}{CNN_{+_{S_\text{i}(M)}} + CNN_{-_{S_\text{i}(M)}}}}{\frac{CNN_{-_{S_\text{i}(M)}}}{CNN_{+_{S_\text{i}(M)}} + CNN_{-_{S_\text{i}(M)}}}} \nonumber \\
                                                           &= \frac{CNN_{+_{S_\text{i}(M)}}}{CNN_{-_{S_\text{i}(M)}}},
\label{equation12}
\end{align}
where $CNN_{+_{S_\text{i}(M)}}$ denotes the number of predictors $S_\text{i}$ formed by a positive link with node $M$ where each predictor includes only node $M$ and excludes the links $(A, M)$ and $(B, M)$. Likewise, $CNN_{-_{S_\text{i}(M)}}$ represents the number of predictors $S_\text{i}$ formed by a negative link with node $M$ in which each predictor includes only node $M$ and excludes the links $(A, M)$ and $(B, M)$. Similar to Eq. (\ref{equation11}), we also add 1 to both the numerator and denominator of Eq. (\ref{equation12}), resulting in
\begin{equation}
\tilde{R}_{S_\text{i}(M)}^{\text{GSMNB-CN}} = \frac{CNN_{+_{S_\text{i}(M)}}+1}{CNN_{-_{S_\text{i}(M)}}+1}.
\label{equation13}
\end{equation}

\subsection{Generalized multiple motifs-based Na\"ive Bayes model}
\label{section34}

Since Eq. (\ref{equation9}) is restricted to assessing the predictive performance of a single predictor, it cannot be directly applied to evaluate the combined predictive power of multiple predictors, which is crucial for understanding their joint contributions to sign prediction. For example, when predictors $S_\text{1}$, $S_\text{2}$, and $S_\text{5}$ are used together for sign prediction (Fig. \ref{Fig:func}A), Eq. (\ref{equation9}) provides no means of quantifying their collective effectiveness. This limitation highlights the need for a theoretical extension that generalizes the evaluation to multiple predictors.

Here, we explore two approaches to quantify the combined contribution of multiple predictors for sign prediction. The first approach leverages Na\"ive Bayes theory, introducing a Generalized Multiple Motifs-based Na\"ive Bayes (GMMNB) model. The second approach adopts a machine learning framework, named FGMNB, where the scores of the nine predictors described in Section \ref{section33} are represented as a 9-dimensional feature vector and input into a classifier to assess their combined effect.

 For the GMMNB method, when nine predictors, $S_\text{1}$ to $S_\text{9}$, are employed for sign prediction, the posterior probability of $(A, B)$ being positive or negative can be calculated as
\begin{align}
&P(+_{AB}|\{S_\text{1}(A,B),S_\text{2}(A,B),..., S_\text{9}(A,B)\})  \nonumber \\
                 &= \frac{P(+_{AB}) \times P(\{S_\text{1}(A,B),S_\text{2}(A,B),..., S_\text{9}(A,B)\}|+_{AB})}{P(\{S_\text{1}(A,B),S_\text{2}(A,B),..., S_\text{9}(A,B)\})} \nonumber \\      
                 &= \frac{P(+_{AB}) \times P(S_\text{1}(A,B)|+_{AB}) \times ... \times P(S_\text{9}(A,B)|+_{AB})}{P(\{S_\text{1}(A,B),S_\text{2}(A,B),..., S_\text{9}(A,B)\})},
\label{equation14}
\end{align}

\begin{align}
&P(-_{AB}|\{S_\text{1}(A,B),S_\text{2}(A,B),..., S_\text{9}(A,B)\})  \nonumber \\
                 &= \frac{P(-_{AB}) \times P(\{S_\text{1}(A,B),S_\text{2}(A,B),..., S_\text{9}(A,B)\}|-_{AB})}{P(\{S_\text{1}(A,B),S_\text{2}(A,B),..., S_\text{9}(A,B)\})} \nonumber \\      
                 &= \frac{P(-_{AB}) \times P(S_\text{1}(A,B)|-_{AB}) \times ... \times P(S_\text{9}(A,B)|-_{AB})}{P(\{S_\text{1}(A,B),S_\text{2}(A,B),..., S_\text{9}(A,B)\})}.
\label{equation15}
\end{align}
Similarly, using Maximum A Posteriori estimation \citep{greig1989exact}, the likelihood score for the sign of the target link (A, B) is computed as 
\begin{align}
r'_{AB} &=\frac{P(+_{AB}|\{S_\text{1}(A,B),S_\text{2}(A,B),..., S_\text{9}(A,B)\})}{P(-_{AB}|\{S_\text{1}(A,B),S_\text{2}(A,B),..., S_\text{9}(A,B)\})} \nonumber \\
            &= \frac{P(+_{AB})}{P(-_{AB})} \times \frac{P(S_\text{1}(A,B)|+_{AB})}{P(S_\text{1}(A,B)|-_{AB})} \times ... \nonumber\\
            & \times \frac{P(S_\text{9}(A,B)|+_{AB})}{P(S_\text{9}(A,B)|-_{AB})}\nonumber \\
            &= \frac{P(+_{AB})}{P(-_{AB})} \times \prod_{M \in S_\text{1}(A,B)} \frac{P(-_{AB})}{P(+_{AB})} \times \frac{P(+_{AB}|M)}{P(-_{AB}|M)}  \nonumber\\
            & \times ... \times \prod_{T \in S_\text{9}(A,B)} \frac{P(-_{AB})}{P(+_{AB})} \times \frac{P(+_{AB}|T)}{P(-_{AB}|T)}.
\label{equation16}
\end{align}
Similarly, let $a = \frac{P(-_{AB})}{P(+_{AB})}=f_-/f_+$, Eq. (\ref{equation16}) can be rewritten as
\begin{equation}
r'_{AB} = a^{-1} \prod_{M \in S_\text{1}(A,B)}a {R_{S_\text{1}(M)}} \times ... \times \prod_{T \in S_\text{9}(A,B)}a {R_{S_\text{9}(T)}}.
\label{equation17}
\end{equation}
Taking the logarithm of Eq. (\ref{equation17}), the new likelihood score for the sign of the target link $(A, B)$ can be calculated as
\begin{align}
\tilde{r}'_{AB} &= \log( \prod_{M \in S_\text{1}(A,B)}a {R_{S_\text{1}(M)}} \times ... \times \prod_{T \in S_\text{9}(A,B)}a {R_{S_\text{9}(T)}}) \nonumber \\
                     &= \sum_{\text{i}=1}^{9}|S_\text{i}(A,B)|\log{a}  + \sum_{\text{i}=1}^{9}(\sum_{M \in S_\text{i}(A,B)}\log{R_{S_\text{i}(M)}}).
\label{equation18}
\end{align}
Finally, by substituting Eq. (\ref{equation11}) or Eq. (\ref{equation13}) into Eq. (\ref{equation18}), we can calculate the likelihood score for the sign of the target link $(A, B)$ under multiple motifs.

\subsection{Time complexity analysis}
\label{section35}
Let $|N|$ and $|L|$ represent the number of nodes and links in a signed network, respectively. The computational complexity of motif counting depends on the motif size and the traversal process. For a given predicted link $(A, B)$, computing the number of 3-node motifs involving $(A, B)$ requires iterating over all nodes in the network, resulting in a time complexity of $O(|N|)$. Similarly, computing the number of 4-node motifs involving $(A, B)$ necessitates traversing all links, leading to a time complexity of $O(|L|)$. The GSMNB-CN and GSMNB-CL models introduce a role function, which requires an additional layer of iteration over all links. Consequently, their time complexities increase to $O(|N||L|)$ for 3-node motifs and $O(|L|^2)$ for 4-node motifs. 

\subsection{XGBoost classifier}
\label{section36}
XGBoost has emerged as a leading machine learning approach for tackling complex classification problems \citep{chen2016xgboost}. Its core strength lies in constructing ensembles of decision trees in a way that progressively reduces errors from earlier models, enabling it to capture intricate and highly non-linear relationships among features. This makes XGBoost particularly effective for modeling the nuanced structural and relational patterns that characterize signed networks. 

A further advantage of XGBoost is its capacity to manage model complexity through an integrated regularization scheme. By incorporating mechanisms that penalize overly complex models, it reduces the risk of overfitting and enhances generalization to unseen data. In addition, users can adjust a wide range of hyperparameters, such as learning rate, tree depth, and minimum child weight, allowing the model to be carefully adapted to the specific characteristics of sign prediction tasks. 

In the context of sign prediction, XGBoost has proven valuable in exploiting diverse sources of information, such as local structural configurations, higher-order motifs, and node-level attributes, to determine whether links are positive or negative \citep{liu2020sign}. Beyond its predictive accuracy, it also offers interpretability by providing measures of feature importance, which can reveal the structural features that most strongly influence sign outcomes. This combination of adaptability, interpretability, and strong performance has made XGBoost a reliable choice for advancing research in signed network analysis.

\section{Experiments and analysis}
\label{section4}

\subsection{Network data}
\label{section41}

To effectively evaluate the prediction performance of our proposed methods, we consider four signed social networks where links are explicitly positive or negative. {\bf BitcoinAlpha} and {\bf BitcoinOTC} \citep{kumar2016edge,kumar2018rev2} are who-trusts-whom networks representing interactions among users trading Bitcoin on their respective platforms. On these platforms, users rate others on a scale from -10 (total distrust) to +10 (total trust), in increments of 1. The sign of a user's rating determines the sign of the links between two users, while the absolute value reflects the link's weight, representing the strength of trust or distrust. These networks are particularly valuable for studying trust dynamics in financial ecosystems where pseudonymity is prevalent.

\begin{table}[htbp]
\centering
\caption{The basic statistics of four empirical signed networks. $|N|$ is the number of nodes, and $|L|$ denotes the number of links in a signed network. $f_+$ and $f_-$ are the fraction of positive and negative signs in a signed network, respectively.}
\begin{tabular}{lcccc}
\toprule
Networks & $|N|$ & $|L|$ & $f_+$ & $f_-$ \\
\midrule
BitcoinAlpha & 3,783 & 14,124 & 91.60\% & 8.40\% \\
BitcoinOTC & 5,881 & 21,492 & 86.42\% & 13.58\% \\
Wiki-RfA & 11,221 & 171,761 & 77.43\% & 22.57\% \\
Slashdot & 82,140 & 500,481 & 76.36\% & 23.64\% \\
\bottomrule
\end{tabular}
\label{table:datasets}
\end{table}

{\bf Wiki-RfA} (Request for Adminship ) \citep{west2014exploiting} captures the community-driven process by which Wikipedia members elect administrators. During this process, a candidate (or another member) submits a request, and community members respond by voting to support, remain neutral, or oppose. These votes form a directed, signed network in which each node represents a Wikipedia member, positive links indicate supporting votes, and negative links indicate opposition. The dataset provides valuable insight into decision-making, community dynamics, and reputation management in a collaborative environment. 

{\bf Slashdot} \citep{leskovec2010signed} is a technology-focused news website where users can label others as either ``friends'' or ``foes''. This tagging system forms a directed, signed network in which nodes represent users and links denote positive (friend) or negative (foe) relationships. The dataset, collected in February 2009, is often used to study social balance, opinion dynamics, and the interplay between positive and negative relationships in online communities.

Here, we focus exclusively on the sign information of links, disregarding their direction and weight, in the four empirical signed networks described. To convert a directed signed network into an undirected one, bidirectional and unidirectional negative links are treated uniformly as undirected negative links, while bidirectional and unidirectional positive links are treated as undirected positive links. Bidirectional links that include both positive and negative signs represent a semantically contradictory and unstable relationship, and therefore cannot be meaningfully classified as either mutual trust or mutual hostility. Following established preprocessing practices \citep{liu2020sign,ma2020snegan}, such inconsistent bidirectional links are treated as unreliable and are removed from the network in this study. The statistics of the resulting undirected signed networks are summarized in Table \ref{table:datasets}.

\subsection{Experimental setup}
\label{section42}
Sign prediction is actually treated as a binary classification problem, with performance evaluated based on the distinction between positive and negative sets. In this paper, we use the likelihood score computed by Eq. (\ref{equation9}) or Eq. (\ref{equation18}) as either a one-dimensional feature or a set of multiple likelihood scores as multi-dimensional features, which are then fed into an XGBoost classifier to perform sign prediction.

Since real networks typically contain more positive signs than negative ones (Table \ref{table:datasets}), we use negative signs as the reference. To ensure the problem is technically testable, 90\% of the negative signs are randomly selected from the original network to form the negative training set, while the remaining 10\% constitute the negative testing set. We adopt balanced positive and negative samples. Positive samples, equal in size to the negative samples, are created by randomly selecting positive signs. For each network, we run the above procedure 100 times, generating 100 network realizations with sampled negative signs and 100 pairs of training and testing sets. In this paper, we report the average prediction performance of 100 network realizations.

\subsection{Baseline methods}
\label{section43}
To illustrate the effectiveness of our proposed methods in sign prediction tasks, we compare them against five baselines.
\begin{itemize}
		
	\item DNE-SBP \citep{shen2018deep}. The Deep Network Embedding with Structural Balance Preservation (DNE-SBP) model is proposed to learn low-dimensional node vector representations while maintaining structural balance in signed networks. The model employs a semi-supervised stacked autoencoder to reconstruct the adjacency matrix of the signed network, with a larger penalty imposed on reconstructing negative links to address the imbalance between positive and negative links. To preserve structural balance, pairwise constraints are introduced, ensuring that positively connected nodes are closer in the embedding space than negatively connected nodes.
	
	\item SGCN \citep{derr2018signed}. SGCN is a neural network model specifically designed for signed networks, where links can have both positive and negative signs. Unlike traditional graph convolutional networks primarily focusing on unsigned networks, SGCN incorporates balance theory to effectively aggregate and propagate information across network layers while addressing the unique challenges of negative links. By leveraging this principled approach, SGCN enhances node representation learning in signed networks, improving performance on tasks such as sign prediction.
	
	\item SiGAT \citep{huang2019signed}. Huang \textit{et al.} introduce the Signed Graph Attention Network (SiGAT), which is an extension of Graph Attention Networks (GATs) designed to handle signed networks, where both positive and negative links exist. SiGAT extends graph attention mechanisms to signed networks by integrating sociological principles and motif-based message passing. This enables the model to learn robust node embeddings and improve performance on tasks like sign prediction.
	
	\item SDGNN \citep{huang2021sdgnn}. SDGNN is an effective and theoretically grounded graph neural network model tailored for signed and directed networks. It is guided by two fundamental sociological theories: status theory and balance theory. By incorporating sociological principles and optimizing for multiple structural properties, SDGNN surpasses previous methods in learning meaningful node embeddings.

    \item SE-SGformer \citep{li2025self} is a self-explainable transformer framework for signed graph neural networks that addresses the challenge of limited interpretability in sign prediction. It employs positional encoding based on signed random walks to achieve greater expressive power than existing SGNN and transformer-based models. By replacing the neural network decoder with a K-nearest (farthest) positive (negative) neighbor mechanism, SE-SGformer delivers both strong predictive performance and clear explanatory insights, surpassing state-of-the-art methods on real-world datasets.

	\end{itemize}

\subsection{Evaluation metrics}
\label{section44}
The most commonly reported metrics in sign prediction are Accuracy and the area under the receiver operating characteristic curve (AUC) \citep{leskovec2010predicting,liu2020sign}. Accuracy is a performance metric used to evaluate the effectiveness of a classification model. It is defined as the proportion of correctly classified instances (both positive and negative) out of the total number of instances. Mathematically, Accuracy is expressed as
\begin{equation}
\text{Accuracy} = \frac{TP+TN}{TP+FN+FP+TN},
\label{equation:acc}
\end{equation}
where $TP$ represents correctly predicted positive instances, $FN$ is positive instances incorrectly classified as negative, $FP$ is negative instances incorrectly classified as positive, and $TN$ represents correctly predicted negative instances.

The AUC can be interpreted as the probability that a randomly chosen positive instance is ranked higher than a randomly chosen negative instance \citep{ran2024maximum,bi2024inconsistency}. A key advantage of AUC is its ability to account for both the true positive rate and the true negative rate, making it a more comprehensive metric for evaluating a classifier's performance, especially in imbalanced datasets. The AUC can be estimated using a sampling method. For each comparison, one instance is randomly selected from the positive testing set and one from the negative testing set, and their predicted probabilities are compared. If out of $n$ comparisons, the positive sample has a higher predicted probability than the negative sample in $n'$ cases, and both have the same predicted probability in $n''$ cases, the AUC can be calculated as
\begin{equation}
\text{AUC} = \frac{n' + 0.5n''}{n}.
\label{equation:auc}
\end{equation}

\subsection{Results on generalized single motif-based Na\"ive Bayes model}
\label{section45}
To assess the effectiveness of GSMNB-CN and GSMNB-CL, we compare their performance with SMNB across four real-world networks. We first compute the score of each predictor $S_\text{i}$ (nine predictors in Fig. \ref{Fig:hm}B) using SMNB, GSMNB-CN, and GSMNB-CL, respectively. The score of each predictor is then used as a one-dimensional feature and fed into an XGBoost classifier for sign prediction.

Tables \ref{table:auc} and \ref{table:acc} present the AUC and Accuracy results across the four datasets. The key findings are as follows. First, SMNB assumes that all links sharing the same neighbor have identical probabilities of being positive or negative, which limits its predictive power. In contrast, GSMNB-CL consistently outperforms SMNB, demonstrating that incorporating common links into the generalized single motif-based Na\"ive bayes model provides a more accurate assessment of a neighboring node's influence. Second, across all datasets, each predictor achieves its highest predictive performance when quantified using GSMNB-CL. This supports our initial hypothesis that the contribution of a common neighbor should be evaluated differently to the sign of different target links depending on their specific connections. Finally, while GSMNB-CL consistently outperforms SMNB, GSMNB-CN often yields lower performance than SMNB. This suggests that the contribution mechanism of the common link plays a crucial role in sign prediction, whereas considering only common nodes without their connecting links may introduce noise or reduce predictive accuracy.

Further analysis based on Na\"ive Bayes theory reveals that on BitcoinAlpha and BitcoinOTC, 3-node predictors outperform 4-node predictors, particularly $S_\text{4}$. Conversely, on Wiki-RfA and Slashdot, 4-node predictors generally yield better performance, especially $S_\text{2}$. These results indicate that incorporating more high-order network information does not always improve prediction accuracy. This finding reinforces our decision to focus on 3-node and 4-node local structural patterns in this study.

\begin{table*}[]
\centering
\caption{Performance comparison of the AUC by each predictor $S_\text{i}$ for sign prediction. The best results under different quantitative methods are marked in bold.} 
\setstretch{1.2}
\resizebox{\textwidth}{!}{
\begin{tabular}{@{}cccccccccccccccc@{}}
\toprule
       & \multicolumn{3}{c}{BitcoinAlpha} &  & \multicolumn{3}{c}{BitcoinOTC}&  & \multicolumn{3}{c}{Wiki-RfA}&  & \multicolumn{3}{c}{Slashdot} \\ 
       \cmidrule(lr){2-4} \cmidrule(l){6-8}  \cmidrule(l){10-12}  \cmidrule(l){14-16} 
Predictors & SMNB  & GSMNB-CN & GSMNB-CL           &  & SMNB    & GSMNB-CN& GSMNB-CL      &  & SMNB     & GSMNB-CN& GSMNB-CL     &  & SMNB    & GSMNB-CN & GSMNB-CL             \\ 
\midrule
$S_\text{1}$ & 0.669 & 0.667& \textbf{0.762}  & & 0.749 & 0.748& \textbf{0.796}  & & 0.690 & 0.675& \textbf{0.738}  & & 0.683 & 0.681& \textbf{0.690}        \\
$S_\text{2}$ & 0.693 & 0.598& \textbf{0.781}  & & 0.801 & 0.758& \textbf{0.862}  & & 0.747 & 0.701& \textbf{0.793}  & & 0.807 & 0.796& \textbf{0.839}        \\ 
$S_\text{3}$ & 0.643 & 0.579& \textbf{0.692}  & & 0.721 & 0.692& \textbf{0.757}  & & 0.694 & 0.649& \textbf{0.744}  & & 0.635 & 0.612& \textbf{0.656}         \\ 
$S_\text{4}$ & 0.771 & 0.760& \textbf{0.814}  & & 0.806 & 0.811& \textbf{0.828}  & & 0.603 & 0.592& \textbf{0.669}  & & 0.572 & 0.569& \textbf{0.587}          \\ 
$S_\text{5}$ & 0.761 & 0.737& \textbf{0.776}  & & 0.787 & 0.766& \textbf{0.811}  & & 0.717 & 0.666& \textbf{0.739}  & & 0.693 & 0.656 & \textbf{0.714}        \\ 
$S_\text{6}$ & 0.680 & 0.666& \textbf{0.732}  & & 0.684 & 0.675& \textbf{0.715}  & & 0.659 & 0.597& \textbf{0.694}  & & 0.616 & 0.593& \textbf{0.628}         \\ 
$S_\text{7}$ & 0.518 & 0.540& \textbf{0.546}  & & 0.535 & 0.540& \textbf{0.567}  & & 0.539 & 0.548& \textbf{0.553}  & & 0.516 & \textbf{0.524}& 0.521          \\ 
$S_\text{8}$ & \textbf{0.666} & 0.661& \textbf{0.666}  & & 0.699 & 0.687& \textbf{0.701}  & & 0.726 & 0.719& \textbf{0.729}  & & 0.606 & \textbf{0.607} & \textbf{0.607}         \\ 
$S_\text{9}$ & 0.685 & 0.688& \textbf{0.691}  & & 0.753 & 0.746& \textbf{0.762}  & & 0.733 & 0.730& \textbf{0.740}  & & \textbf{0.658} & 0.654& \textbf{0.658}        \\ 
\bottomrule
\end{tabular}
}
\label{table:auc}
\end{table*}

\begin{table*}[]
\centering
\caption{Performance comparison of the Accuracy by each predictor $S_\text{i}$ for sign prediction. The best results under different quantitative methods are marked in bold.} 
\setstretch{1.2}
\resizebox{\textwidth}{!}{
\begin{tabular}{@{}cccccccccccccccc@{}}
\toprule
       & \multicolumn{3}{c}{BitcoinAlpha} &  & \multicolumn{3}{c}{BitcoinOTC}&  & \multicolumn{3}{c}{Wiki-RfA}&  & \multicolumn{3}{c}{Slashdot} \\ 
       \cmidrule(lr){2-4} \cmidrule(l){6-8}  \cmidrule(l){10-12}  \cmidrule(l){14-16} 
Predictors & SMNB  & GSMNB-CN& GSMNB-CL            &  & SMNB    & GSMNB-CN& GSMNB-CL      &  & SMNB     & GSMNB-CN & GSMNB-CL    &  & SMNB     & GSMNB-CN& GSMNB-CL             \\ 
\midrule
$S_\text{1}$ & 0.624 & 0.620& \textbf{0.701}  & & 0.723 & 0.721& \textbf{0.748}  & & 0.654 & 0.638& \textbf{0.697}  & & 0.672 & 0.670& \textbf{0.675}        \\
$S_\text{2}$ & 0.633 & 0.569& \textbf{0.717}  & & 0.728 & 0.707& \textbf{0.784}  & & 0.677 & 0.648& \textbf{0.718}  & & 0.722 & 0.709& \textbf{0.755}        \\ 
$S_\text{3}$ & 0.601 & 0.556& \textbf{0.631}  & & 0.649 & 0.630& \textbf{0.671}  & & 0.643 & 0.612& \textbf{0.685}  & & 0.613 & 0.588& \textbf{0.628}         \\ 
$S_\text{4}$ & 0.727 & 0.731& \textbf{0.771}  & & 0.765 & 0.770& \textbf{0.794}  & & 0.585 & 0.581& \textbf{0.630}  & & 0.564 & 0.562& \textbf{0.572}          \\ 
$S_\text{5}$ & 0.707 & 0.699& \textbf{0.727}  & & 0.734 & 0.728& \textbf{0.754}  & & 0.672 & 0.623& \textbf{0.688}  & & 0.656 & 0.631& \textbf{0.667}         \\ 
$S_\text{6}$ & 0.638 & 0.639& \textbf{0.654}  & & 0.648 & 0.640& \textbf{0.666}  & & 0.621 & 0.578& \textbf{0.648}  & & 0.580 & 0.561& \textbf{0.587}         \\ 
$S_\text{7}$ & 0.523 & 0.533& \textbf{0.536}  & & 0.535 & 0.537& \textbf{0.552}  & & 0.543 & 0.545& \textbf{0.548}  & & 0.514 & 0.516& \textbf{0.517}         \\ 
$S_\text{8}$ & 0.646 & 0.652& \textbf{0.654}  & & 0.681 & 0.673& \textbf{0.685}  & & 0.697 & 0.697& \textbf{0.698}  & & 0.603 & 0.605& \textbf{0.610}          \\ 
$S_\text{9}$ & 0.672 & \textbf{0.674}& 0.667  & & 0.728 & 0.722& \textbf{0.736}  & & \textbf{0.710} & 0.709& 0.708  & & \textbf{0.652} & 0.650& \textbf{0.652}        \\ 
\bottomrule
\end{tabular}
}
\label{table:acc}
\end{table*}

From Tables \ref{table:auc} and \ref{table:acc}, it is clear that the predictive performance of the nine motifs differs substantially. To explain this variation, we introduce a quantitative metric called \textit{Motif Coverage}. For a given motif type $S$, let $\mathcal{M}_S(l)$ denote the set of all motif instances of type $S$ that include link $l \in L$. We define the indicator function as follows:  

\begin{equation}
I_S(l) =
\begin{cases}
1, & \text{if } \mathcal{M}_S(l) \neq \emptyset, \\
0, & \text{otherwise}.
\end{cases}
\end{equation}
The Motif Coverage of motif $S$ is then defined as:  
\begin{equation}
\text{Motif Coverage} = \frac{1}{|L|} \sum_{l \in L} I_S(l).
\end{equation}

\begin{figure}[htbp]
	\centering
	\includegraphics[width=1.0\textwidth]{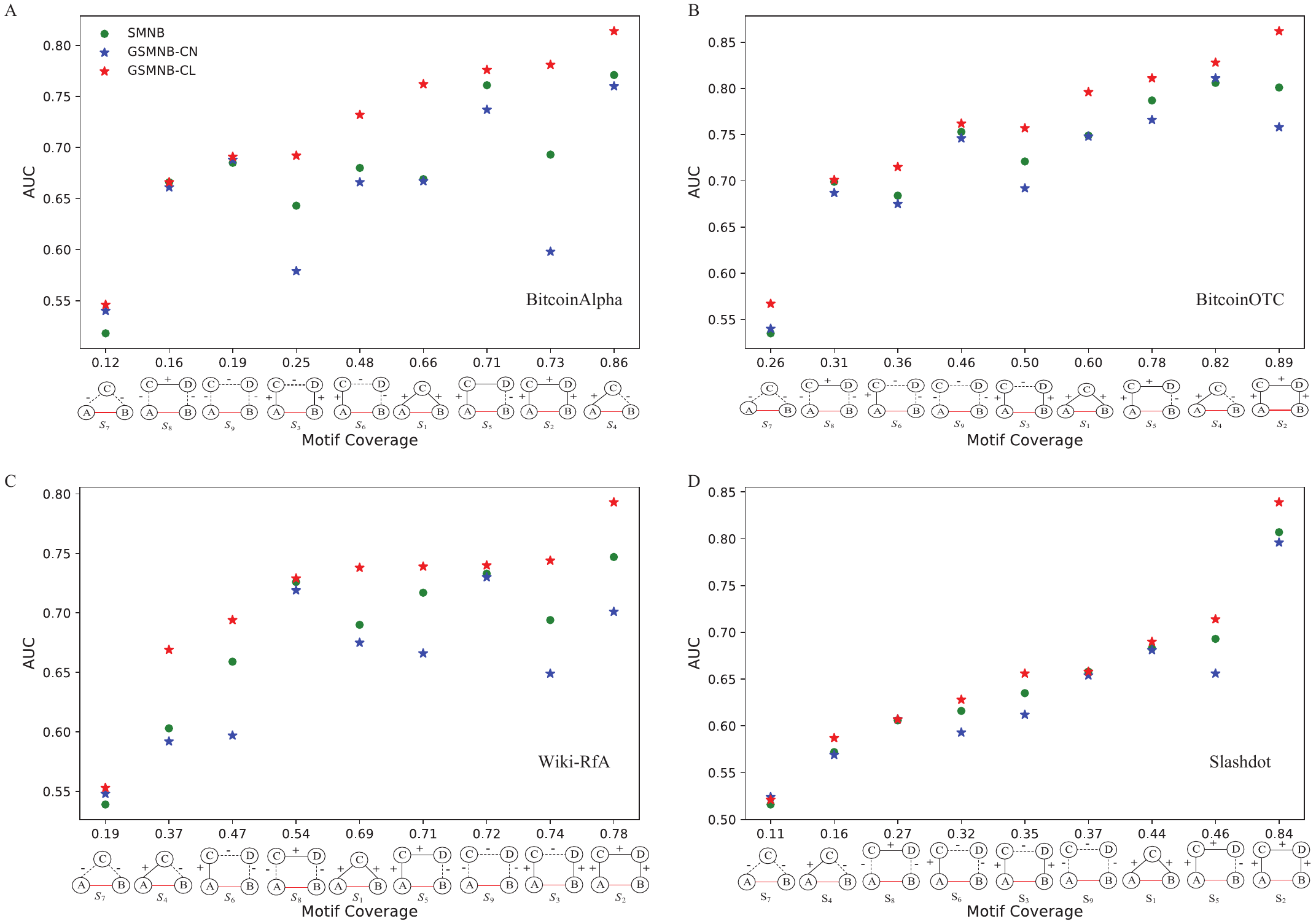}
	\caption{The Relationship Between Motif Coverage and Algorithm Performance (AUC). The sign prediction performance of different algorithms, particularly GSMNB-CL, shows a positive correlation with motif coverage. Specifically, as the motif coverage of a given motif type increases within the network, the corresponding AUC also tends to increase.}
	\label{Fig:fig4}
\end{figure}

Intuitively, Motif Coverage measures the fraction of links in the network that participate in at least one instance of motif $S$. For example, the coverage of motif $S_\text{1}$ is obtained by counting the number of links that, together with their neighbors, form motif $S_\text{1}$, and then normalizing by the total number of links in the network.

As shown in Fig. \ref{Fig:fig4}, we compute the coverage of each motif across four networks. The results show that motif coverage is a key determinant of predictive performance. In BitcoinAlpha and BitcoinOTC, AUC increases strongly with motif coverage, with GSMNB-CL consistently achieving the best results. Although the trend is less consistent in Wiki-RfA, high-coverage motifs such as $S_\text{2}$ still yield superior accuracy. A similar pattern emerges in Slashdot, where greater motif coverage leads to higher AUC, and GSMNB-CL shows the largest improvements.

Overall, motif coverage serves as a reliable indicator of predictive effectiveness in sign prediction. High-coverage motifs contribute disproportionately to performance gains, with GSMNB-CL benefiting most, underscoring the importance of local motif structures in this task.

\subsection{Results on generalized multiple motifs-based Na\"ive Bayes model}
\label{section46}
The results in Section \ref{section45} demonstrate that the contribution mechanism of the common link substantially enhances sign prediction performance. Building on this insight, we extend contribution mechanisms based on common links to multiple motifs by substituting Eq. (\ref{equation11}) into Eq. (\ref{equation18}), thereby implementing the GMMNB and FGMNB methods. Here, we evaluate their effectiveness through experiments on four signed social networks, comparing them against five state-of-the-art network embedding methods. 

\begin{figure}[htbp]
	\centering
	\includegraphics[width=1.0\textwidth]{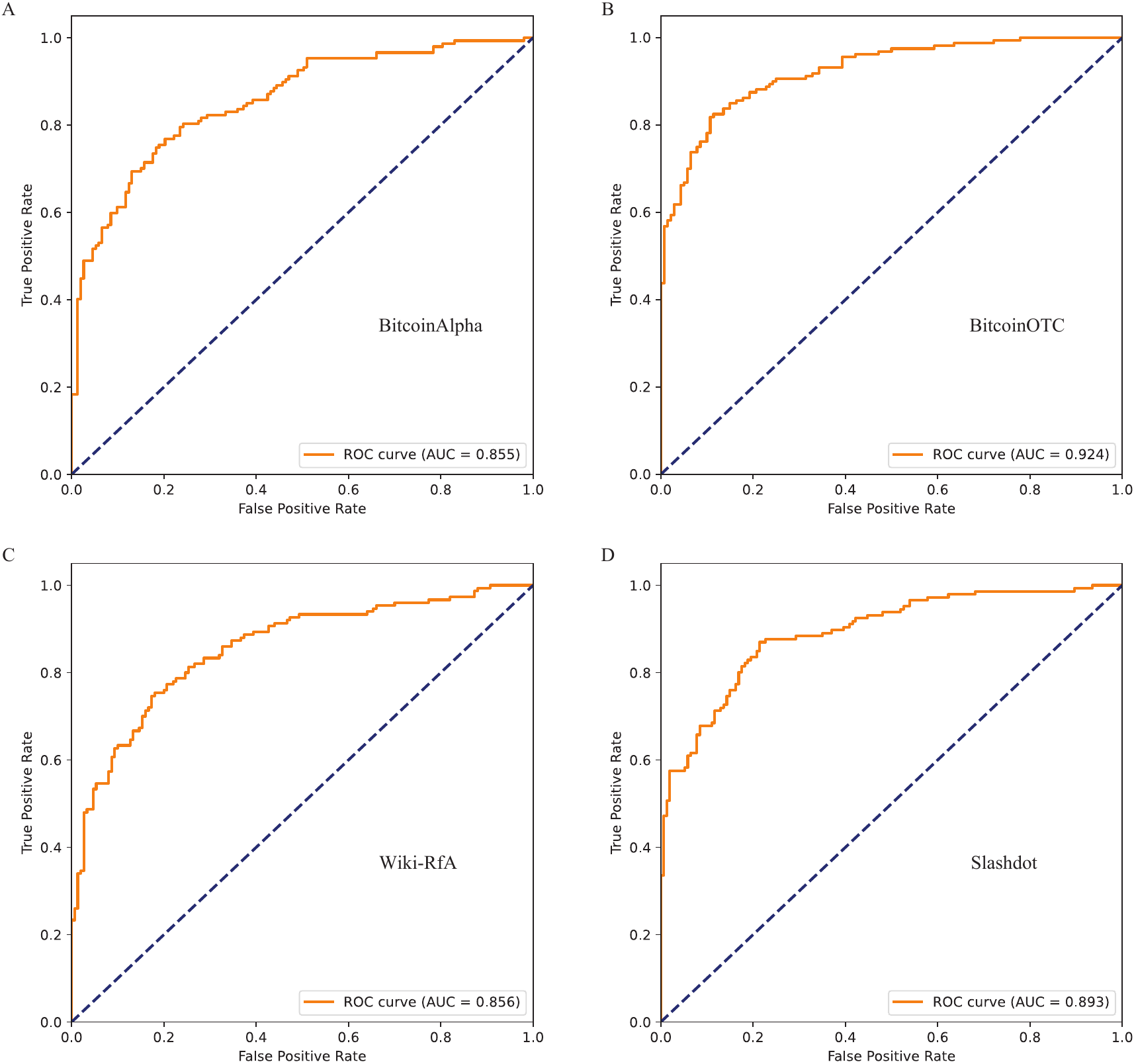}
	\caption{The Receiver Operating Characteristic (ROC) curves for FGMNB. The ROC curves demonstrate a pronounced rise above the diagonal baseline, reflecting the FGMNB’s strong discriminative capability.}
	\label{Fig:fig5}
\end{figure}

Tables \ref{table:compauc} and \ref{table:compacc} compare the performance of multiple methods on four real-world signed networks using AUC and Accuracy as evaluation metrics. As shown in Table \ref{table:compauc}, FGMNB achieves the highest AUC on three datasets, highlighting its superior predictive performance. Consistently, its ROC curves rise well above the diagonal baseline, further demonstrating strong discriminative ability (Fig. \ref{Fig:fig5}). GMMNB also performs competitively, generally ranking second and consistently outperforming embedding-based baselines.

In contrast, embedding-based baselines deliver weaker and less consistent performance. Although DNE-SBP performs relatively well among them, its AUC and Accuracy remain well below those of GMMNB and FGMNB. The embedding–based methods such as SGCN, SiGAT, and SDGNN perform particularly poorly on Wiki-RfA and Slashdot, suggesting that their embeddings may fail to capture sign-related structural dependencies in certain network contexts. This limitation is likely exacerbated by the influence of link direction on embedding-based models.

Moreover, SE-SGformer, based on the graph transformer, achieves the highest AUC on Slashdot and the highest accuracy on both Wiki-RfA and Slashdot, confirming the utility of graph transformer techniques in sign prediction. Nevertheless, FGMNB outperforms SE-SGformer in AUC across all three networks, indicating that local structural information is more critical for sign prediction. Unlike SE-SGformer, which leverages global information, FGMNB achieves superior performance using only local patterns, demonstrating the effectiveness of motif-based modeling.

\begin{table}[htbp]
\centering
\caption{Performance comparison of the AUC across different methods for sign prediction. Here, we report the average performance and standard deviation. The best results under different methods are marked in bold.}
\begin{tabular}{lcccc}
\toprule 
Methods & BitcoinAlpha & BitcoinOTC & Wiki-RfA & Slashdot \\
\midrule
DNE-SBP & 0.722$\pm$0.02 & 0.836$\pm$0.01 & 0.760$\pm$0.03 & 0.812$\pm$0.02 \\
SGCN & 0.826$\pm$0.01 & 0.811$\pm$0.02 & 0.580$\pm$0.05 & 0.656$\pm$0.04 \\
SiGAT & 0.792$\pm$0.02 & 0.831$\pm$0.02 & 0.553$\pm$0.04 & 0.681$\pm$0.03 \\
SDGNN & 0.806$\pm$0.02 & 0.820$\pm$0.01 & 0.582$\pm$0.03 & 0.710$\pm$0.04 \\
SE-SGformer & 0.703$\pm$0.04 & 0.721$\pm$0.04 & 0.800$\pm$0.10 & \textbf{0.974$\pm$0.01} \\
GMMNB & 0.802$\pm$0.02 & 0.903$\pm$0.03 & 0.819$\pm$0.04 & 0.866$\pm$0.02 \\
FGMNB & \textbf{0.851$\pm$0.02} & \textbf{0.920$\pm$0.01} & \textbf{0.853$\pm$0.02} & 0.894$\pm$0.01 \\
\bottomrule
\end{tabular}
\label{table:compauc}
\end{table}

\begin{table}[htbp]
\centering
\caption{Performance comparison of the Accuracy across different methods for sign prediction. Here, we report the average performance and standard deviation. The best results under different methods are marked in bold.}
\begin{tabular}{lcccc}
\toprule 
Methods & BitcoinAlpha & BitcoinOTC & Wiki-RfA & Slashdot \\
\midrule
DNE-SBP & 0.678$\pm$0.02 & 0.733$\pm$0.02 & 0.651$\pm$0.03 & 0.711$\pm$0.03 \\
SGCN & 0.762$\pm$0.02 & 0.742$\pm$0.02 & 0.540$\pm$0.04 & 0.580$\pm$0.02 \\
SiGAT & 0.724$\pm$0.02 & 0.757$\pm$0.04 & 0.532$\pm$0.02 & 0.635$\pm$0.02 \\
SDGNN & 0.743$\pm$0.02 & 0.744$\pm$0.02 & 0.560$\pm$0.03 & 0.660$\pm$0.03 \\
SE-SGformer & 0.708$\pm$0.03 & 0.726$\pm$0.02 & \textbf{0.833$\pm$0.02} & \textbf{0.976$\pm$0.01} \\
GMMNB & 0.758$\pm$0.02 & 0.822$\pm$0.02 & 0.725$\pm$0.02 & 0.784$\pm$0.01 \\
FGMNB & \textbf{0.784$\pm$0.02} & \textbf{0.845$\pm$0.02} & 0.773$\pm$0.03 & 0.814$\pm$0.01 \\
\bottomrule
\end{tabular}
\label{table:compacc}
\end{table}

The superior performance of GMMNB and FGMNB highlights the importance of motif-based features in sign prediction. GMMNB, which employs a linear combination of nine motif structures using Na\"ive Bayes theory, already provides a significant improvement over embedding-based baselines, demonstrating that explicitly integrating motif-based information enhances predictive power. FGMNB further refines this approach by treating the nine motifs as independent feature dimensions and applying a machine-learning model, leading to even greater predictive accuracy. This result may highlight the advantage of feature-driven learning over simple linear combinations.

These findings reveal a key limitation of conventional network-embedding approaches for signed networks: while embeddings capture general structural properties, they may overlook local motif structures essential for understanding signed interactions. Although SE-SGformer occasionally outperforms on individual datasets, the consistent advantage of leveraging local motif structures underscores the effectiveness and robustness of FGMNB for sign prediction tasks. Overall, by directly modeling these motif structures using Na\"ive Bayes theory, GMMNB and FGMNB provide a more interpretable and effective solution for sign prediction, with FGMNB emerging as the most robust and accurate method.

\section{Conclusion and Discussion}
\label{section5}
To summarize, we address the limitation of the SMNB method by introducing two enhanced approaches, GSMNB-CL and GSMNB-CN, to more accurately capture the heterogeneous influence of a neighboring node on a link's sign. Our experiments on four large-scale signed social networks reveal that GSMNB-CL consistently outperforms SMNB, underscoring the positive impact of the contribution mechanism of the common link in sign prediction. By contrast, GSMNB-CN often underperforms SMNB, indicating that focusing solely on a neighbor node---while excluding its shared links---can diminish predictive power.

We further extend the contribution mechanism of the common link to handle multiple motifs through two frameworks: GMMNB, which uses a Na\"ive Bayes model to linearly combine the contributions of nine different motifs, and FGMNB, which treats each motif as an independent feature dimension in a machine learning model. Analysis of four real-world signed networks demonstrates the superiority of motif-based methods for sign prediction. FGMNB consistently achieves the highest AUC and strong ROC performance, while GMMNB ranks second, outperforming embedding-based baselines that often fail to capture local sign-dependent structures. Although SE-SGformer performs well on certain datasets, FGMNB’s consistent advantage underscores the importance of local motifs. By treating motifs as independent features, FGMNB surpasses GMMNB’s linear approach, offering a more interpretable, robust, and accurate solution than conventional embedding-based methods.

Our findings also suggest that higher-order motifs (4-node motifs) do not always guarantee better performance than simpler 3-node structures. Specifically, 3-node motifs are more effective on the BitcoinAlpha and BitcoinOTC datasets, while 4-node motifs excel on Wiki-RfA and Slashdot. This underscores the complexity of signed networks, where the choice of local structural patterns can substantially influence predictive outcomes. Nonetheless, our results confirm that motif-based approaches by Na\"ive Bayes model are generally more effective and interpretable than embedding-based methods in capturing the nuanced interactions that underlie signed networks.

Despite the strong performance of the proposed motif-based approaches by Na\"ive Bayes model, our study also has certain limitations that open avenues for future research. Our study focuses on static signed networks, where the structure and link signs remain unchanged during analysis. However, real-world signed networks often evolve, with nodes forming and dissolving connections dynamically \citep{li2017fundamental,li2020evolution}. Future work could extend our framework to dynamic signed networks, incorporating temporal higher-order motifs to capture evolving patterns in sign formation. Additionally, while our methods are computationally efficient compared to deep learning-based approaches, the calculation of motif-based probabilities still requires processing local structures, which can be computationally intensive for very large-scale networks \citep{slota2014complex,lin2016network}. Future research could focus on developing more scalable motif-based algorithms, such as approximating motif contributions using sampling techniques or leveraging graph sparsity to reduce computational overhead.

Although this approach is clear and intuitive, it assumes that the contributions of common neighbors are independent. A promising direction for future work is to relax this assumption and model the dependencies among common neighbors \citep{jiang2008novel}, thereby enabling a more accurate quantification of their interactions. Another potential avenue is to integrate GSMNB-CL and GSMNB-CN into a unified framework and investigate whether such integration can yield performance gains over either individual model. Finally, incorporating the proposed framework into a graph neural network architecture could further enhance its representational capacity and predictive performance.

\section*{CRediT authorship contribution statement}
\textbf{Yijun Ran:} Conceptualization, Methodology, Investigation, Visualization, Writing - original draft, Writing - review \& editing, Funding acquisition. \textbf{Si-Yuan Liu:} Conceptualization, Methodology, Software. \textbf{Junjie Huang:} Conceptualization, Methodology, Software, Funding acquisition. \textbf{Tao Jia:} Conceptualization, Methodology, Supervision, Formal analysis, Writing - review \& editing, Funding acquisition. \textbf{Xiao-Ke Xu:} Conceptualization, Methodology, Supervision, Formal analysis, Writing - review \& editing, Funding acquisition.

\section*{Declaration of competing interest}
The authors declare that they have no known competing financial interests or personal relationships that could have appeared to influence the work reported in this paper.

\section*{Data availability}
The datasets used in this study are available at \href{http://snap.stanford.edu/data/index.html}{SNAP}.

\section*{Acknowledgments}
This work is supported by the National Natural Science Foundation of China (Nos. 62403062, 72374173, 62173065, and 62402398), the Beijing Natural Science Foundation (No. 4242040), the University Innovation Research Group of Chongqing (No. CXQT21005), the Fundamental Research Funds for the Central Universities (Nos. 124330008, SWU-XDJH202303, and 123330009), and the Postdoctoral Fellowship Program of CPSF (No. GZC20230281). This study's high-performance computations are supported by the Interdisciplinary Intelligence Supercomputer Center at Beijing Normal University, Zhuhai, China.

\end{document}